\documentclass[%
 aip,
 amsmath,amssymb,
 reprint,%
]{revtex4-1}

\usepackage{graphicx}
\usepackage{dcolumn}
\usepackage{bm}
\usepackage[english]{babel}
\usepackage[utf8]{inputenc}
\usepackage[T1]{fontenc}
\usepackage{mathptmx}
\usepackage{color}
\usepackage{siunitx}
\usepackage{color}
\usepackage{bm}
\usepackage{multirow}

\usepackage{ulem}
\usepackage{ifthen}
\definecolor{orange}{rgb}{1,0.5,0}
\definecolor{darkgreen}{rgb}{0,0.4,0.1}

\newcommand*\bfb{{\bf b}}
\newcommand*\bfr{{\bf r}}
\newcommand*\bff{{\bf f}}
\newcommand*\bfF{{\bf F}}
\newcommand*\bfk{{\bf k}}
\newcommand*\bfP{{\bf P}}
\newcommand*\bfR{{\bf R}}
\newcommand*\bfq{{\bf q}}

\newcommand*\bfz{{\bf z}}
\newcommand*\fe{{\mathcal{F}}}
\begin{document}

\preprint{AIP/123-QED}

\title{{Computing three-dimensional densities from force densities improves statistical efficiency} }

\author{Samuel W. Coles}
\affiliation
{Sorbonne Universit\'{e}, CNRS, Physicochimie des \'{e}lectrolytes et nanosyst\`{e}mes interfaciaux, UMR PHENIX, F-75005, Paris, France}

\author{Daniel Borgis}
\affiliation
{PASTEUR, D\'{e}partement de chimie, \'{E}cole Normale Sup\'{e}rieure, PSL University, Sorbonne Universit\'{e}, CNRS, 75005 Paris, France}
\affiliation{Maison de la Simulation, CEA, CNRS, Universit\'{e} Paris-Sud, UVSQ, Universit\'{e} Paris-Saclay, 91191 Gif-sur-Yvette, France}

\author{Rodolphe Vuilleumier}
\affiliation{PASTEUR, D\'{e}partement de chimie, \'{E}cole Normale Sup\'{e}rieure, PSL University, Sorbonne Universit\'{e}, CNRS, 75005 Paris, France}

\author{Benjamin Rotenberg}
\email{benjamin.rotenberg@sorbonne-universite.fr}
\affiliation
{Sorbonne Universit\'{e}, CNRS, Physicochimie des \'{e}lectrolytes et nanosyst\`{e}mes interfaciaux, UMR PHENIX, F-75005, Paris, France}

\date{\today}

\begin{abstract}
The extraction of inhomogeneous 3-dimensional densities around
tagged solutes from molecular simulations 
is known to have a very high computational cost because this is
traditionally performed by collecting histograms, with each discrete voxel in
three-dimensional space needing to be visited significantly. This paper presents
an extension of a previous methodology for the extraction of 3D solvent number
densities with a reduced variance principle [Borgis \textit{et al.}, Mol. Phys.
\textbf{2013} , 111, 3486-3492] to other 3D densities such as charge and
polarization densities. The approach is also generalized to cover molecular
solvents with structures described using rigid geometrical constraints, which
include in particular popular water models such as SPC/E and TIPnP class of
models. The noise reduction is illustrated for the microscopic hydration
structure of a small molecule, in various simulation conditions, and for a protein. The method has large applicability to simulations of solvation in many fields, for example around biomolecules, nanoparticles, or within porous materials.
\end{abstract}

\maketitle

\section{Introduction}

Understanding and predicting the microscopic water structure around biological macromolecule is of primary importance in structural
biology as well as in drug discovery. In the latter case, the desolvation cost
of the water molecules with highest affinities around a binding site, {\it
i.e.}, the free-energy cost of removing them in order to make ligand binding
possible is widely believed to be the main source of the overall binding free
energy. Therefore, mapping the locations and thermodynamic properties of water
molecules close to protein binding sites from the analysis of the local
(orientation-dependent) solvent density offers rich physical insights. Such
analysis is at the heart of the WaterMap approach\cite{Abel2007,Abel2008}, or of
the grid inhomogeneous solvation theory (GIST) of Gilson and collaborators, who
use the solvent densities measured from explicit solvent simulations to estimate
localized solvation entropies, energies, and free energies\cite{Nguyen2012}.
Using such approaches, they were able to decipher the functional role of water
molecules according to their location. In more global structural studies, the
knowledge of the water density around biomolecules is fundamental to reproduce
or predict small angle neutron scattering (SANS) or X-Ray scattering (SAXS)
spectra\cite{Nguyen2014,Marchi2016} or the location of the ``crystallographic''
water molecules in X-ray diffraction
spectra\cite{altan_learning_2018,wall_biomolecular_2019}. It is also a
key quantity when identifying the hydrophobic and hydrophilic sites of proteins
and understanding the contribution of water to their folded structure or their association. Since modeling biomolecules in solution involves very large systems, often with easily tens or hundreds of thousands water molecules to be simulated, the accumulation of statistics to get the 3-dimensional (3D) water density with sufficient accuracy entails a huge computational cost. 

Extraction and analysis of 3D densities also has applications to a wide array of
problems involving nanomaterials in solution or solid-liquid interfaces. In
electrochemistry, for instance, current atomic force microscopy techniques allow
for the measurement of interfacial charge densities which can be
rationalized by comparison to simulations\,\cite{klausen_mapping_2016}.
Recently a two-dimensional structure of ionic liquids has been observed at
electrodes by means of atomic force
microscopy\cite{josesegura_adsorbed_2013,elbourne_nanostructure_2015} ; the
propagation of this structure into the bulk could be more thoroughly explored
using molecular dynamics simulations via the use of 3-dimensional
densities expanding on previous work using interfacial 2-dimensional density\,\cite{docampo-alvarez_molecular_2016,kornyshev_three-dimensional_2014,merlet_electric_2014}. Another particular point of interest is the study of supercapacitors and the accurate determination of the 3D charge density in and around the porous carbon electrodes\cite{jeanmairet_molecular_2019,merlet_molecular_2012,merlet_highly_2013,kondrat_effect_2012,simoncelli_blue_2018}, in order to understand and, if possible, optimize the capacitance. We note finally that predicting the hydration structure around complex molecular objects such as proteins is the playground for liquid-state statistical mechanics approaches such as 3D-RISM\,\cite{Yoshida2009,Stumpe2011}, or molecular density functional theory\cite{Ding2017,jeanmairet_molecular_2019,Zhao2011,Jeanmairet2013}, and having at one's disposal precise solvent densities that can serve as references for those approaches is critical for development and validation.

For all those instances, the accurate evaluation of 3D solvent densities from
molecular simulations represents a key numerical issue. This creates an impetus
for the development of any method which would allow the acquisition to be done
as efficiently as possible and the simulation time minimised. This paper will
present a new method to do just this. The natural way to evaluate those
quantities, which consists of accumulating presence probability histograms in 3D
space, suffers from high statistical noise, with each voxel in 3D space needing
to be significantly visited by individual solvent molecules. In addition,
the process is mathematically ill-defined because the variance of this estimator
tends to infinity as $1/\Delta v$ when the voxel volume $\Delta v$ tends to
zero. 

In previous work\cite{borgis_computation_2013}, we introduced a method to
compute pair distribution functions and 3-dimensional densities with a
reduced variance principle, by noting that the ensemble averages defining these
observables can be re-expressed in a statistically better-behaved form after
integrating by parts with respect to the atomic coordinates. For the 3D-density,
the method consists in sampling the force density instead of the density itself,
and reconstructing the latter in a subsequent step. This approach was recovered
from a different perspective by de las Heras and
Schmidt\cite{de_las_heras_better_2018}, who used as a starting point the fact
that the force density is in fact equal to the gradient of the density (up to a
factor $k_BT$, the thermal energy). 
Finally, the force sampling approach has also been reported recently
by Purohit, Schultz and Kofke as an instance of ``mapped averaging'', 
a general framework to compute properties from molecular 
simulation\cite{purohit_force-sampling_2019,schultz_alternatives_2019}.

Here we expand the theory developed in
Ref.\citenum{borgis_computation_2013}, which is extended in two important
directions. On the one hand, we show how to apply it in the practically relevant
case of molecular models involving constraints, in particular rigid molecules
such as the popular SPC/E water model. On the other, we discuss how to go beyond
number densities and compute physically important properties such as charge and
polarization densities. This force density route to compute number, charge, and
polarization densities is applied to a series of test cases in order to quantify
the reduction of the variance with respect to the standard histograms
collection. We study the 3D hydration structure around a tagged water molecule
in SPC/E water under different simulation conditions, as well as that of a
prototypical protein, specifically lysozyme.

\section{Theory}

\subsection{Number density from force density}

We first recall here a few results obtained in
Ref.\citenum{borgis_computation_2013}. For a simple fluid of $N$ identical
particles submitted to an external potential field, the inhomogeneous number
density is defined from their positions $\bfr_i$ as:
\begin{align}
\rho( \bfr ) &= \left\langle \sum_{i=1}^N \delta( \bfr_i - \bfr ) \right\rangle
\label{eq:rhodef}
\end{align}
\noindent 
where $\delta$ is the Dirac delta function, and $\left\langle \dots \right\rangle$ denotes an average in the canonical ensemble. 
Differentiating this expression relates the gradient of the density to
the force density as:
\begin{align}
\nabla \rho( \bfr ) &= 
\beta \bfF( \bfr ) =
\beta \left\langle \sum_{i=1}^N \delta( \bfr_i - \bfr ) \bff_i \right\rangle
\label{eq:gradrhonoconstraints}
\end{align} 
\noindent 
where $\bff_i$ is the force acting on particle $i$ and $\beta=1/k_BT$.

A key step is of course to determine the density $\rho(\bfr)$ from its gradient,
sampled from a trajectory as the force density via Eq.~\ref{eq:gradrhonoconstraints}. 
The integration constant can be determined from
the average density $\rho_0$. Choosing the antiderivative
with an average value of zero amounts to working with
the excess density $\Delta \rho(\bfr) = \rho(\bfr) - \rho_0$.
Introducing the Fourier transform of a function
$g(\bfk)=\int g(\bfr) e^{-i\bfk\cdot\bfr}{\rm d}\bfr $
(with corresponding inverse transform
$g(\bfr)=\frac{1}{(2\pi)^3}\int g(\bfk) e^{+i\bfk\cdot\bfr}{\rm d}\bfk$),
Eq.~\ref{eq:gradrhonoconstraints} can be rewritten
in Fourier space as $i\bfk\Delta \rho(\bfk)=\beta\bfF(\bfk)$.
Taking the dot product with $i\bfk$, which amounts to taking the divergence
of both vector fields, then gives the inverse of the gradient operator as:  
\begin{align}\label{eq:rho-by-FFT} 
\Delta\rho(\bfk)&= -\frac{i\beta}{k^2} \bfk\cdot\bfF(\bfk)
\; .
\end{align}
Since $i\bfk/k^2$ is the Fourier transform
of $\nabla(\frac{1}{4\pi r})=-\frac{\bfr}{4\pi r^3}$, this last result is
equivalent to a convolution in real space: 
\begin{align}
\Delta\rho(\bfr)&= \frac{\beta}{4\pi} \displaystyle \int 
\frac{ \bfr'-\bfr }{ |\bfr'-\bfr|^3 } 
\cdot \bfF(\bfr') \ {\rm d}\bfr'
\; .
\end{align}
This was the core formula derived by another route in
Ref.\citenum{borgis_computation_2013}. The convolution is more conveniently
performed numerically in reciprocal space using Eq.~\ref{eq:rho-by-FFT} and
Fast Fourier transforms (FFT), even though of course any other
numerical method to inverse the gradient can be used.

We now propose to generalize those formulas to the more realistic case of a molecular liquid composed of rigid molecules described by distributed atomic sites and geometrical constraints applying between the sites. To this end, we stick to the density gradient route adopted above and follow the general approach of Ciccotti \textit{et al.} to compute potentials of mean force from molecular dynamics simulations in the presence of constraints\cite{ciccotti_blue_2005}; this will be applied later on to a specific set of collective, geometrical variables.

\subsection{Number density in the presence of constraints}

\subsubsection{Constraints, collective variables and potential of mean force}

We first recall the main finding of Ref.~\citenum{ciccotti_blue_2005}, in which the authors derived (with slightly different notations) an expression of the mean force along generic collective variables $\bfq(\bfr^N)=\left\{ q_1(\bfr^N), \dots, q_K(\bfr^N) \right\}$, where $K$ is the number of such variables and $\bfr^N =\left\{ \bfr_1, \dots, \bfr_N \right\}$ the set of all atomic coordinates, subject to $M$ molecular constraints written as $\sigma(\bfr^N)=0$, where $\sigma(\bfr^N)=\left\{ \sigma_1(\bfr^N), \dots, \sigma_M(\bfr^N) \right\}$. The most common examples of such constraints include fixed bond distances, which can be written as $|\bfr_j-\bfr_j'|^2 - d^2 = 0$, or fixed angles between bonds. The free energy $\fe$ associated with the set of collective variables is given by,
\begin{widetext}
\begin{align}
e^{-\beta \fe(\bfz)}  := \frac{1}{Z_\sigma}
\displaystyle 
\int e^{-\beta U(\bfr^N)} \prod_{k=1}^K \delta( q_k(\bfr^N) - z_k )
\prod_{m=1}^M \delta(\sigma_m(\bfr^N))\ {\rm d}\bfr^N 
\; ,
\label{eq:freeen}
\end{align}
\end{widetext}
\noindent 
where $\bfz=\left\{ z_1, \dots, z_K \right\}$ is a specific value of the vectorial collective variable, $U(\bfr^N)$ is the potential energy of the microscopic configuration and
\begin{align}
Z_\sigma = 
\displaystyle 
\int e^{-\beta U(\bfr^N)} \prod_{m=1}^M \delta(\sigma_m(\bfr^N))\ {\rm d}\bfr^N 
\; .
\label{eq:partfunc}
\end{align}

The main result of Ciccotti {\it et al.} is then that the gradient of the free energy $\fe(\bfz)$, {\it i.e.} the mean force associated with a change in the value of the collective variables, can be expressed as
\begin{align}
\frac{\partial \mathcal{F}}{\partial z_k}
&= \left\langle
\bfb_k(\bfr^N)\cdot\nabla U - k_BT \nabla\cdot\bfb_k(\bfr^N)
\right\rangle_{ \bfq(\bfr^N) = \bfz\, , \, \sigma(\bfr^N)=0}
\; ,
\label{eq:pmf}
\end{align}
\noindent 
where the conditional average of a function $f$ is defined as,
\begin{widetext}
\begin{align}
\left\langle f\right\rangle_{ \bfq(\bfr^N) = \bfz\, , \, \sigma(\bfr^N)=0}
&:= \frac{
\displaystyle 
\int f(\bfr^N) e^{-\beta U(\bfr^N)} \prod_{k=1}^K \delta( q_k(\bfr^N) - z_k )
\prod_{m=1}^M \delta(\sigma_m(\bfr^N))\ {\rm d}\bfr^N 
}{
\displaystyle 
\int e^{-\beta U(\bfr^N)} \prod_{k=1}^K \delta( q_k(\bfr^N) - z_k )
\prod_{m=1}^M \delta(\sigma_m(\bfr^N))\ {\rm d}\bfr^N 
}
\; ,
\label{eq:condavg}
\end{align}
\end{widetext}

\noindent 
and where the $\bfb_k(\bfr^N)$, for $k=1,\dots,K$ are 3N-dimensional vector fields satisfying,
\begin{align}
\left\{
  \begin{array}{lr}
\bfb_k(\bfr^N) \cdot \nabla \sigma_m(\bfr^N) = 0
\quad \forall k=1,\dots,K \, ; \, \forall m=1,\dots,M
\\
\bfb_k(\bfr^N) \cdot \nabla q_{k'}(\bfr^N) = \delta_{k,k'}
\; , \; {\it i.e.}\ 1 {\rm \ if}\  k=k' {\rm \ and\ 0\ otherwise}.
\end{array}
\right.
\label{eq:bfield}
\end{align}

\noindent
Here we note that the gradients in these equations are with respect to the 3N components of the full set of positions $\bfr^N$. We will now use this result to obtain a workable expression for the number density in the presence of molecular constraints and to extend this beyond the number density.

\subsubsection{Position as a collective variable}

In order to use the above generic expression to compute the local number density, we now consider a very simple choice of $K=3$ collective variables, namely the coordinates of a single particle $i$, $\bfq(\bfr^N)=\bfr_i$, which form a subset of the 3N-dimensional vector $\bfr^N$ (with the indices $3i-2$, $3i-1$ and $3i$). The probability to find this particle at position $\bfr$ is given by $e^{-\beta \fe_i(\bfr)}$, with $\fe_i$ defined by Eq.~\ref{eq:freeen} for the particular choice $\bfq(\bfr^N)=\bfr_i$ and $\bfz=\bfr$. The following argument can be reproduced separately for each particle $i$ to compute the corresponding $\fe_i(\bfr)$, from which we straightforwardly obtain the density $\rho(\bfr)=\sum_{i=1}^N e^{-\beta \fe_i(\bfr)}$ via Eq.~\ref{eq:rhodef}.

To proceed further, we limit ourselves to the standard case where only distance constraints are present. Note that these constraints may or may not involve the selected particle $i$, but that the canonical averages are performed taking all constraints into account. We can build the vector fields satisfying Eq.~\ref{eq:bfield} by computing first the 3N-dimensional gradients of the constraints
$\sigma_m(\bfr^N)=|\bfr_j^m-\bfr_{j'}^m|^2-d_m^2$,
where the $m$-th constraint fixes the distance between particles $j$ and $j'$ to $d_m$, and of the collective variables $q_k(\bfr^N)$, specifically $q_1(\bfr^N)=x_i$, $q_2(\bfr^N)=y_i$, and $q_3(\bfr^N)=z_i$, the coordinates of particle $i$.
For the constraints, one obtains,
\begin{widetext}
\begin{align}
\nabla \sigma_m(\bfr^N) = 
( 0, 0, 0, \dots, 
&\underbrace{ 2(x_j-x_{j'}) , 2(y_j-y_{j'}), 2(z_j-z_{j'})}_{ {\rm atom}\ j }, \dots,
\underbrace{ -2(x_j-x_{j'}) , -2(y_j-y_{j'}), -2(z_j-z_{j'})}_{ {\rm atom}\ j' },
\dots,  0, 0, 0 )
\label{eq:gradsigma}
\end{align} 
\end{widetext}
\noindent
while for the collective variables, the gradients read:
\begin{align}\label{eq:gradq}
\nabla q_1(\bfr^N) &= ( 0, 0, 0, \dots, 1, 0, 0, \dots, 0, 0, 0 ) \nonumber \\
\nabla q_2(\bfr^N) &= ( 0, 0, 0, \dots, 0, 1, 0, \dots, 0, 0, 0 ) \nonumber \\
\nabla q_3(\bfr^N) &= ( 0, 0, 0, \dots, \underbrace{0, 0, 1}_{ {\rm atom}\ i }, \dots,  0, 0, 0 )
\end{align} 

We can now construct a solution to Eq.~\ref{eq:bfield} as follows. For $k=$1, 2
and 3, we consider the 3N-dimensional ${\bf b}_k(\bfr^N)$ vector with 0
everywhere except at the $(3i-3+k)$ indices (corresponding to the $x$, $y$ and
$z$ coordinates of particle $i$) and the $(3j-3+k)$ indices for all particles
$j$ involved in a constraint with particle $i$
(with indices $j^{i}_{1}$ to $j^{i}_{m_i}$, where $m_i$
is the number of such constraints). For these indices, the value of
the corresponding coordinate of ${\bf b}_k(\bfr^N)$ is taken equal to 1.
This can be written explicitly as:
\begin{widetext}
\begin{align}
\label{eq:bsolution}
{\bf b}_1(\bfr^N) &= (  0, 0, 0,\dots, 1, 0, 0,                               \dots, 1, 0, 0,                                \dots, 1, 0, 0,                                 \dots, 0, 0, 0 ) \nonumber \\
{\bf b}_2(\bfr^N) &= ( 0, 0, 0, \dots, 0, 1, 0,                               \dots, 0, 1, 0,                                \dots, 0, 1, 0,                                 \dots, 0, 0, 0 ) \nonumber \\
{\bf b}_3(\bfr^N) &= ( 0, 0, 0, \dots, \underbrace{0, 0, 1}_{ {\rm atom}\ j^{i}_{1}},\dots, \underbrace{0, 0, 1}_{ {\rm atom}\ i}, \dots, \underbrace{0, 0, 1}_{{\rm atom}\ j^{i}_{m_i}}, \dots,  0, 0, 0 )
\end{align} 
\end{widetext}
It is easy to check by taking the dot products with the gradients in
Eqs.~\ref{eq:gradsigma} and~\ref{eq:gradq} that the ${\bf b}_k(\bfr^N)$ so
defined satisfy Eq.~\ref{eq:bfield}. We can then insert this solution into
Eq.~\ref{eq:pmf}. Since the above solution does not depend explicitly on the
positions $\bfr^N$,  $\nabla\cdot\bfb_k(\bfr^N)=0$
and the second term in the ensemble average
vanishes. As for the first, we note that the negative gradient of the potential
energy $U$ reads:
\begin{align}\label{eq:gradU}
-\nabla U &= ( f_{1x},f_{1y},f_{1z},\dots,f_{Nx},f_{Ny},f_{Nz}  ) 
\end{align} 
with $f_{i\alpha}$ the components of the force acting on particle $i$.
Therefore, with the above definition in Eq.~\ref{eq:bsolution},
the dot product with $\bfb_k(\bfr^N)$ in Eq.~\ref{eq:pmf}
reduces, up to a minus sign, to the $x$, $y$ and
$z$ components (for $k=1$, 2 and 3) of {\it the sum of forces acting on particle $i$
and all particles participating in a constraint with $i$:}
\begin{align}
\bff_i^* &= \bff_i + \sum_{ m=1 }^{ m_i }
\bff_{j^{i}_m}
\label{eq:totforce} 
\end{align}

These results allow us to obtain a workable expression for the gradient of the
density $\nabla\rho(\bfr)$. For each particle $i$, we now combine
Eqs.~\ref{eq:freeen} and~\ref{eq:pmf} to compute the gradient of $e^{-\beta
\fe_i(\bfr)}$, which reads:
\begin{align}
\nabla e^{-\beta \fe_i(\bfr)} &= -\beta e^{-\beta \fe_i(\bfr)} \nabla \fe_i(\bfr)
= \beta e^{-\beta \fe_i(\bfr)} 
\left\langle  \bff_i^* \right\rangle_{ \bfr_i = \bfr\, , \, \sigma(\bfr^N)=0}
\label{eq:gradexpfi}
\end{align} 
\noindent 
The conditional average (restricted to particle $i$ being at position $\bfr$) of the total force in Eq.~\ref{eq:gradexpfi} can be rewritten as an unconditional average by using Eqs.~\ref{eq:freeen}, \ref{eq:partfunc} and~\ref{eq:condavg}. The final result can be obtained by summing over all particles $i$ which yields
\begin{align}
\nabla \rho( \bfr ) &= 
\beta \bfF( \bfr ) =
\beta \left\langle \sum_{i=1}^N \delta( \bfr_i - \bfr ) \bff_i^*
\right\rangle_{\sigma(\bfr^N)=0}
\; .
\label{eq:gradrho}
\end{align} 
\noindent
where $\bff_i^*$ is defined in Eq.~\ref{eq:totforce} and the average is still taken over configurations satisfying the constraints. This non-trivial result generalizes Eq.~\ref{eq:gradrhonoconstraints} in the presence of constraints of fixed distances, which is of great practical importance in molecular simulations of realistic systems -- in particular with rigid water models, as will be illustrated below. The final step is of course to determine the density $\rho(\bfr)$ from the force density as discussed in the previous section.

\subsection{Beyond number densities: charge and polarization densities}

The derivation of Eq.~\ref{eq:pmf} (hence of Eq.~\ref{eq:gradrho}) relies on an integration by parts with respects to the atomic coordinates $\bfr^N$. As a result, it can be readily extended to any combination of the type,
\begin{align}
A(\bfr) &= \left\langle \sum_{i=1}^N \delta( \bfr_i - \bfr ) a_i \right\rangle
\; ,
\label{eq:aiINTRO}
\end{align}
\noindent where the microscopic property $a_i$ does {\it not} depend on the coordinates $\bfr^N$. A physically relevant example is the charge density,
\begin{align}
\rho_q(\bfr) &= \left\langle \sum_{i=1}^N \delta( \bfr_i - \bfr ) q_i \right\rangle
\; ,
\end{align}
with $q_i$ the partial charge of atom $i$. The derivation of the previous section can then be followed step by step to obtain the final result,
\begin{align}
\nabla A( \bfr ) &= 
\beta \left\langle \sum_{i=1}^N \delta( \bfr_i - \bfr ) a_i \bff_i^*
\right\rangle_{\sigma(\bfr^N)=0}
\; .
\label{eq:gradrhoa}
\end{align} 
\noindent
This quantity can be sampled from the trajectory; the density $A(\bfr)$ is then obtained from its gradient as described above for the number density.

Another microscopic density of particular interest is the polarization density,
which depends on the orientation of the molecules. The electric polarization
$\bfP(\bfr)$ can in principle be determined from the knowledge of the charge
density $\rho_q(\bfr)$, but it is also usual to sample its dipolar component
from the dipoles of molecules. In fact, for 
point dipoles (including,
for example, the Stockmayer model), the 3 components of $\bfP(\bfr)$ can be
obtained from their gradient via Eq.~\ref{eq:gradrhoa}, using the $x$, $y$ and
$z$ components of the dipole
$\bm{\mu}_i$ of each atom as the microscopic property $a_i$.

For polar molecules with explicit charged sites, Eq.~\ref{eq:gradrhoa} cannot be
used directly, because the molecular orientation depends on the atomic
coordinates $\bfr^N$. In the case of rigid molecules (including the SPC/E or
TIP$n$P family of water models), it is nevertheless possible to use a different
set of variables to describe each microscopic configuration, by introducing the
position $\bfR_{I}$ of the center of mass (c.o.m.) and the orientation
$\bm{\Omega}_I$, described by 3 Euler angles ($\theta_I,\phi_I,\psi_I$)
 of each rigid body $I=1,\dots,N_r$. The integrals over coordinates
$\bfr^N$ in the presence of the constraints $\sigma(\bfr^N)$ are then replaced
by unconstrained integrals over the set of c.o.m. positions and orientations
$(\bfR^{N_r},\bm{\Omega}^{N_r})$, and the dipole densities defined as:
\begin{align}
\bfP(\bfr) &= 
\left\langle \sum_{I=1}^{N_r}\delta( \bfR_{I} - \bfr )\bm{\mu}_I(\bm{\Omega}_I) \right\rangle
\nonumber \\
&= \frac{1}{Z} \int {\rm d} \bfR^{N_r} {\rm d} \bm{\Omega}^{N_r} e^{-\beta U(\bfR^{N_r},\bm{\Omega}^{N_r})} 
          \sum_{I=1}^{N_r}\delta( \bfR_{I} - \bfr )  \bm{\mu}_I(\bm{\Omega}_I)
\; ,
\end{align}
\noindent
where the dipole $\bm{\mu}_I$ of each rigid molecule depends on its
orientation $\bm{\Omega}_I$ but not on the position of its c.o.m. $\bfR_{I}$,
and the partition function is $Z=\int {\rm d} \bfR^{N_r} {\rm d} \bm{\Omega}^{N_r}
e^{-\beta U(\bfR^{N_r},\bm{\Omega}^{N_r})}$.
Taking the gradient of $P_\alpha(\bfr)$, 
the $\alpha\in\{x,y,z\}$ components of the polarization density,
with respect to $\bfr$, we can
replace on the right hand-side each term $\nabla_{\bfr}\delta(\bfR_{I} - \bfr )$
in the sum by $-\nabla_{\bfR_{I}}\delta(\bfR_{I} - \bfr )$. Integration by parts
with respect to $\bfR_{I}$ then provides:
\begin{align}
\nabla_\bfr \bfP(\bfr) &= 
\frac{1}{Z} \int {\rm d} \bfR^{N_r} {\rm d} \bm{\Omega}^{N_r} 
          \sum_{I=1}^{N_r}\delta( \bfR_{I} - \bfr )  \bm{\mu}_{I,\alpha}(\bm{\Omega}_I) 
\nabla_{\bfR_{I}}e^{-\beta U(\bfR^{N_r},\bm{\Omega}^{N_r})} 
\nonumber\\ &= 
\frac{\beta}{Z} \int {\rm d} \bfR^{N_r} {\rm d} \bm{\Omega}^{N_r}  e^{-\beta U(\bfR^{N_r},\bm{\Omega}^{N_r})}
\times \nonumber \\
& \, \hskip 1cm  \sum_{I=1}^{N_r}\delta( \bfR_{I} - \bfr )  \bm{\mu}_{I,\alpha}(\bm{\Omega}_I) 
          \left[ -\nabla_{\bfR_{I}} U(\bfR^{N_r},\bm{\Omega}^{N_r}) \right]
\; ,
\label{eq:gradpol}
\end{align}
where we have also used:
$\nabla_{\bfR_{I}}e^{-\beta U(\bfR^{N_r},\bm{\Omega}^{N_r})}
=-\beta e^{-\beta U(\bfR^{N_r},\bm{\Omega}^{N_r})}\nabla_{\bfR_{I}} U(\bfR^{N_r},\bm{\Omega}^{N_r})$.
The last term in square brackets in Eq.~\ref{eq:gradpol} 
is the opposite of the gradient of the energy with respect 
to the position of the c.o.m. of the rigid body $I$, \textit{i.e.} the
(total) force $\bff_I^*$ acting on it.
As a result, the gradients of the components of the polarization are equal to
\begin{align}
\nabla P_\alpha(\bfr) &= 
\beta \left\langle \sum_{I=1}^{N_r} \delta( \bfR_{I} - \bfr ) 
\mu_{I,\alpha} \bff_I^* \right\rangle
\; .
\end{align}
\noindent
The components of $\bfP(\bfr)$ can finally be determined from their gradients as previously.

\section{Systems and Methods}

\subsection{Generation of the Model Systems}

This paper will make use of two model systems; One based on a small box of pure water, and the other of a single lysozyme protein solvated in water. In all cases the solvent used is water modeled using the SPC/E model\,\cite{berendsen_missing_1987}.

\subsubsection{Bulk Water}

Most of the analysis performed in this work uses simulations of bulk water. The
system consists of a 40\,\AA\ $\times$ 40\,\AA\ $\times$ 40\,\AA\ box containing
2113 water molecules, which corresponds to equilibrium density at 300\,K. The
simulations were performed using the Gromacs 2018 molecular dynamics
software\,\cite{abraham_gromacs:_2015}. As this system is invariant by
translation and rotation, average number and charges densities as well as the
average polarization density are known exactly (the average number density is
uniform while the average charge and polarization densities are equal to zero).
This simulation will then serve as a benchmark to study the standard deviations
in the different approaches to evaluate these densities. However, in order to
investigate the structure of water around a given water molecules, we also
simulated a variant of this system in which a water molecule is frozen in space
at the center of the box with oxygen atom centered at
$(20.00{\,},{\,}22.08{\,},{\,}20.00)$ and the molecule is aligned such that the
$z$ axis sits normal to the molecular plane, the dipole of the water molecule is
parallel to the $y$ axis, and the $x$ axis runs along the vector between the two
hydrogen atoms in the constrained water molecule. Initial configurations were
generated using the packmol algorithm\,\cite{martinez_packmol:_2009} followed by
steepest descent energy minimization. This system was then annealed from 300\,K to 500\,K and back to 300\,K over the course of 2\,ns, with a 1\,fs timestep. This was followed by a 1\,ns equilibration run at 300\,K. Finally a 320\,ns production run was performed with a snapshot taken every 100\,fs.

A subsequent 4\,ns production run, with snapshots taken every 10\,fs (though not all these snapshots will be used in every analysis) was performed in order to study the effect of the method on shorter trajectories. Finally a further 4\,ns production run, with snapshots taken every 10\,fs, without the constrained central water molecule was performed. For all simulations a cut-off of 12\,\AA\ is used for both electrostatic and Lennard-Jones potentials. Long range Coulombic interactions are calculated using the Particle Mesh Ewald method (with a maximum error of $\num{1e-6}$)\cite{darden_particle_1993}. All simulations were run in the NVT ensemble using a a Nos\'{e}-Hoover thermostat with used with a time constant of 0.1\,ps \,\cite{posch_canonical_1986}.

\subsubsection{Lysozyme in water}

In order to apply the methodology to more complex systems, we simulate the
solvation of a lysozyme protein (PDB reference 1AKI) by SPC/E water. The system
is adapted from the Gromacs 2018 tutorial written by Justin
Lekmul\,\cite{lemkul_proteins_2018}. The protein is modeled using the OPLS-AA forcefield with input files generated using the pdb2gmx utility within Gromacs 2018\,\cite{jorgensen_development_1996}. The system is modified from the tutorial in that a cubic box is used, and that throughout the simulation the protein atoms are frozen in space. The system consists of the frozen protein solvated by 10644 SPC/E water molecules\,\cite{berendsen_missing_1987}. After steepest descent minimization and equilibration in the NVT (50\,ps), and NPT (100\,ps) ensembles, using the Gromacs native velocity re-scaling thermostat\,\cite{bussi_canonical_2007} and berendsen barostat\,\cite{berendsen_molecular_1984}, we fix the box edge lengths to 69.537\,\AA\ in all dimensions. A 2\,ns production run was then performed in the NVT ensemble using the Gromacs native velocity rescalling thermostat\,\cite{bussi_canonical_2007}. In line with the initial tutorial a 2\,fs time step is used for all simulations. Lennard-Jones, and electrostatic cut-offs are set to be equal to 1\,nm. Long range electrostatics are calculated using the PME solver (with a maximum relative error in energies of $\num{1e-4}$)\,\cite{darden_particle_1993}. Snapshots were taken every 10 steps, or 20\,fs, all of which were used in subsequent analyses.

\subsection{Histogramming MD trajectories}

In order to apply the new method we must generate a cubic grid of force density
from the molecular trajectories. Here we use two schemes to map the continuous
molecular positions onto the discrete grid points; first a conventional 3D
histogram. Second, the inverse of tri-linear interpolation, which we henceforth
will call the triangular kernel method (for reasons that will become apparent). Unlike in the conventional histogram, where all particles in a voxel are placed at its center this second method ascribes a proportion of each particle to each vertex of a voxel in it which it is contained, with the proportion of the particle mapped to each vertex being linearly dependent on the closeness of the particle to the vertex. Both of these methods can be described using a kernel notation where the density is defined as,

\begin{widetext}
\begin{align}
{\rho(i,j,k)} = \dfrac{1}{N_f}\sum_{f=1}^{N_f} \sum_{n=1}^N \frac{K(x_{n,f}-x(i))K(y_{n,f}-y(j))K(z_{n,f}-z(k))}{\delta^3},
\label{eq:methodOfCalculationrho}
\end{align} 
\end{widetext}
\noindent
where $i$, $j$, and $k$ are the grid indices in the $x$, $y$, and $z$
directions, and $x(i)$, $y(j)$, and $z(k)$ are the locations of the grid point
in each dimension for a given index. $x_{n,f}$, $y_{n,f}$, $z_{n,f}$ are the
locations of particle $n$ at a timestep $f$, in each of the grid directions,
$N_f$ is the number of frames,  and $\delta$ is the distance between grid
points. Finally, $K(x_{n,f}-x(i))$ is the kernel, which is defined as,
\begin{align}
K(x_n-x(i))=\left\{
  \begin{array}{lccr}
1 &\operatorname{for}& \left|x_{n,f}-x(i)\right| < \delta/2 
\\
0 &\operatorname{for}& \left|x_{n,f}-x(i)\right|\geq \delta/2
\end{array}
\right.
\label{eq:methodOfCalculationK}
\end{align}
for the box kernel that recovers a conventional histogram and as,
\begin{align}
K(x_n-x(i))=\left\{
  \begin{array}{lccr}
1-\left|x_{n,f}-x(i)\right|/\delta &\operatorname{for}& \left|x_{n,f}-x(i)\right| < \delta 
\\
0 &\operatorname{for}& \left|x_{n,f}-x(i)\right|\geq \delta
\end{array}
\right.
\label{eq:methodOfCalculationKT}
\end{align}
for the triangular kernel, with $K(y_{n,f}-y(j))$, and $K(z_{n,f}-z(k))$ defined
equivalently to $x$ for both kernels. In all cases, the separation should be
calculated with special care given to the system's periodic boundary conditions,
so the distance between the particle and the vertex that is taken is the
shortest possible distance, even if this means taking the distance across the
periodic boundary. Due to the small size of the supports, neither kernel places particle density outside the voxel the particle is located in. The main advantage of the triangular kernel is that it minimizes digitization. The force density can be defined analogously as, 

\begin{widetext}
\begin{align}
{\bfF(i,j,k)} = \dfrac{1}{N_f} \sum_{f=1}^{N_f} \sum_{n=1}^N \frac{K(x_{n,f}-x(i))K(y_{n,f}-y(j))K(z_{n,f}-z(k))\bff_n}{\delta^3}.
\label{eq:methodOfCalculationF}
\end{align} 
\end{widetext}

Once this force density is obtained it can be converted to a number density by using a Fast Fourier Transform to convert $\bfF( \bfr )$ to $\bfF( \bfk )$ and then use of Eq.\,\eqref{eq:rho-by-FFT} and an inverse Fast Fourier Transform to return to real space. In this paper in the interest of speed this step is performed for the average force density, not for each individual timestep.

\subsection{Implementation}

We have implemented the new methodology for the calculation of 3D densities by use of post processing scripts written in the Python programming language. A general script was used to handle Gromacs trajectories using the mdanalysis family of molecular dynamics analysis software\,\cite{gowers_mdanalysis:_2016,michaudagrawal_mdanalysis:_2011}. The generality of this script is such that it could be used for any trajectory format which contains the forces acting on the atoms in the system and can be read by an mdanalysis parser. 
\section{Results and Discussion}

The remainder of the paper presents the results obtained in order to explore the efficacy and relevance of this new force density based methodology. It is split into two main parts: one on the quantification of the relationship between noise and grid spacing, and one looking at the application of the method to study solvation in exemplar simulations. 

\subsection*{Bulk density fluctuations}

\begin{center}
\begin{figure*}[t!]
  \includegraphics[width=\linewidth]{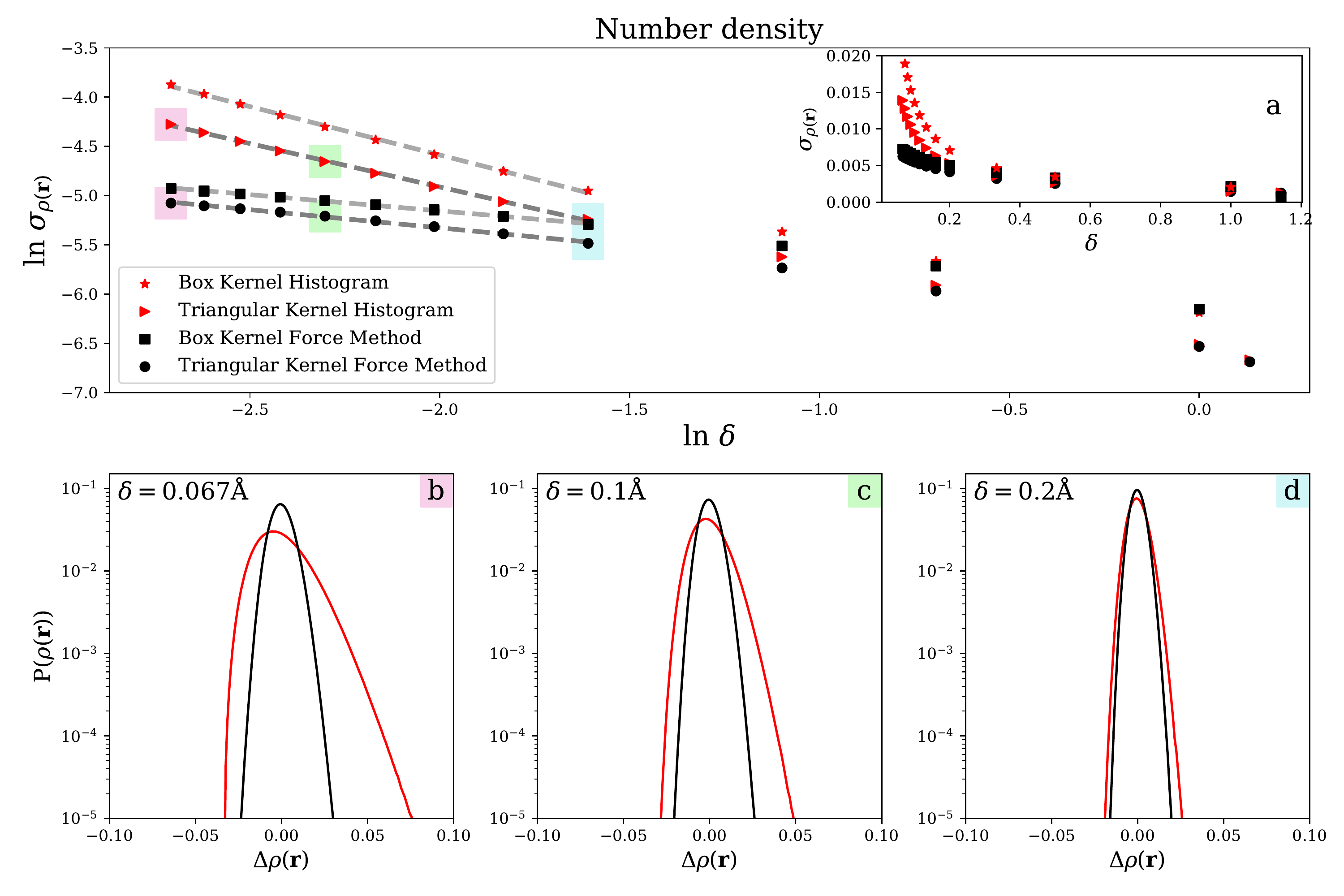}
  \caption{Panel a shows effect of changing $\delta$ on the standard deviation
of the number densities for the histogram method (red stars for box kernel
grids; red triangles for triangular kernel grids), and the force method (black
squares for box kernel grids; black circles for triangular kernel grids), shown
on a log-log scale in the main figure and on a linear scale in the inset. Panels
b through d show the distributions of densities obtained for three delta values
$\delta =$ 0.067\AA, 0.1\AA, and 0.2\AA. The resulting distributions of
$\Delta\rho(\bfr)$ for the histogram method (red) and the force method (black).
The data points on panel a which correspond to the data in panels b through d are shown by colored squares. Dashed lines are linear fits to the data in the region from $\delta=0.2$ to $\delta=0.067$, with the slopes of these lines reported in Table\,1.}
  \label{fig:noisenorm}
\end{figure*}
\end{center}

We start by analyzing the predictions of the density with respect to the
traditional histogram and force methods for the total density in bulk water. We
quantify in particular the effect of the grid spacing for a given number of
configurations (the effect of the latter will be discussed below). In all cases
we employ cubic grids, however we vary the length of the edge of each voxel
(hereon referred to as $\delta$) and compare the effect of this change
between the two methods, and the two different types of kernel. We compare the difference in the standard deviation of the equilibrium number densities obtained for each grid point,
\begin{align}
\sigma_{{\rho(i,j,k)}} =\sqrt{ \dfrac{1}{N_i N_j N_k}\sum_{i=1}^{N_x}\sum_{j=1}^{N_y}\sum_{k=1}^{N_z}\rho(i,j,k)^2-\overline{\rho}^2},
\label{eq:methodOfCalculationSig}
\end{align} 
\noindent
where $i$, $j$, and $k$ are the indices for the bins in the $x$, $y$, and $z$
dimensions, which have $N_x$, $N_y$, and $N_z$ bins. And $\overline{\rho}$ is
the mean density across all the grid points and frames. The trend for the four
combinations of kernel and number density calculation method is shown
in Fig.~\ref{fig:noisenorm}a. The distributions  in
the subsequent three panels are only presented for the triangular kernel.

Looking first to the general trend in standard deviations in
Fig.~\ref{fig:noisenorm}a we observe, as one would expect, that the
standard deviation in number density increases with decreasing $\delta$ for all
four routes. This increase can be effectively approximated by a power law in all
four cases and is thus linear on the log scale. However, for small voxels the
rate of the increase in variance is diminished for the force method relative to
the conventional histogram method, as anticipated from Borgis \textit{et
al.}\,\cite{borgis_computation_2013}. This is demonstrated by the much lower
slope for the linear fits in Fig.~\ref{fig:noisenorm} which are reported in
Table~\ref{table:slopes}.

The probability distributions of number densities in
Fig.~\ref{fig:noisenorm}b through Fig.~\ref{fig:noisenorm}d
(obtained using the triangular kernel) are narrower using the force method when
contrasted to the histogram method for all values of $\delta$ (even though they
converge for large voxels as shown in panel a). There is also a large difference
in the forms of the distributions produced by the two methods, and how
they change with decreasing $\delta$. For $\delta=0.2$\,\AA\ the form of both
distributions is largely Gaussian. For the force method, this remains the same
as $\delta$ decreases with a simple increase in width. In contrast, for the
histogram method, the shape of the distributions changes with decreasing
$\delta$. The peak of the distribution becomes shifted to values of
$\Delta\rho(\mathbf{r})$ below 0, with a fat tail emerging at positive values,
and the distribution becomes increasingly skewed. The asymmetry is due to the
fact that in any one frame only 0.001\%-0.02\% of voxels are occupied,
and as a voxel cannot be sampled fewer than zero times this leads to a warped
distribution of density across the 3D grid. This shows a clear advantage for the
use of the force method when extracting number densities.

\begin{center}
\begin{figure*}[t!]
  \includegraphics[width=\linewidth]{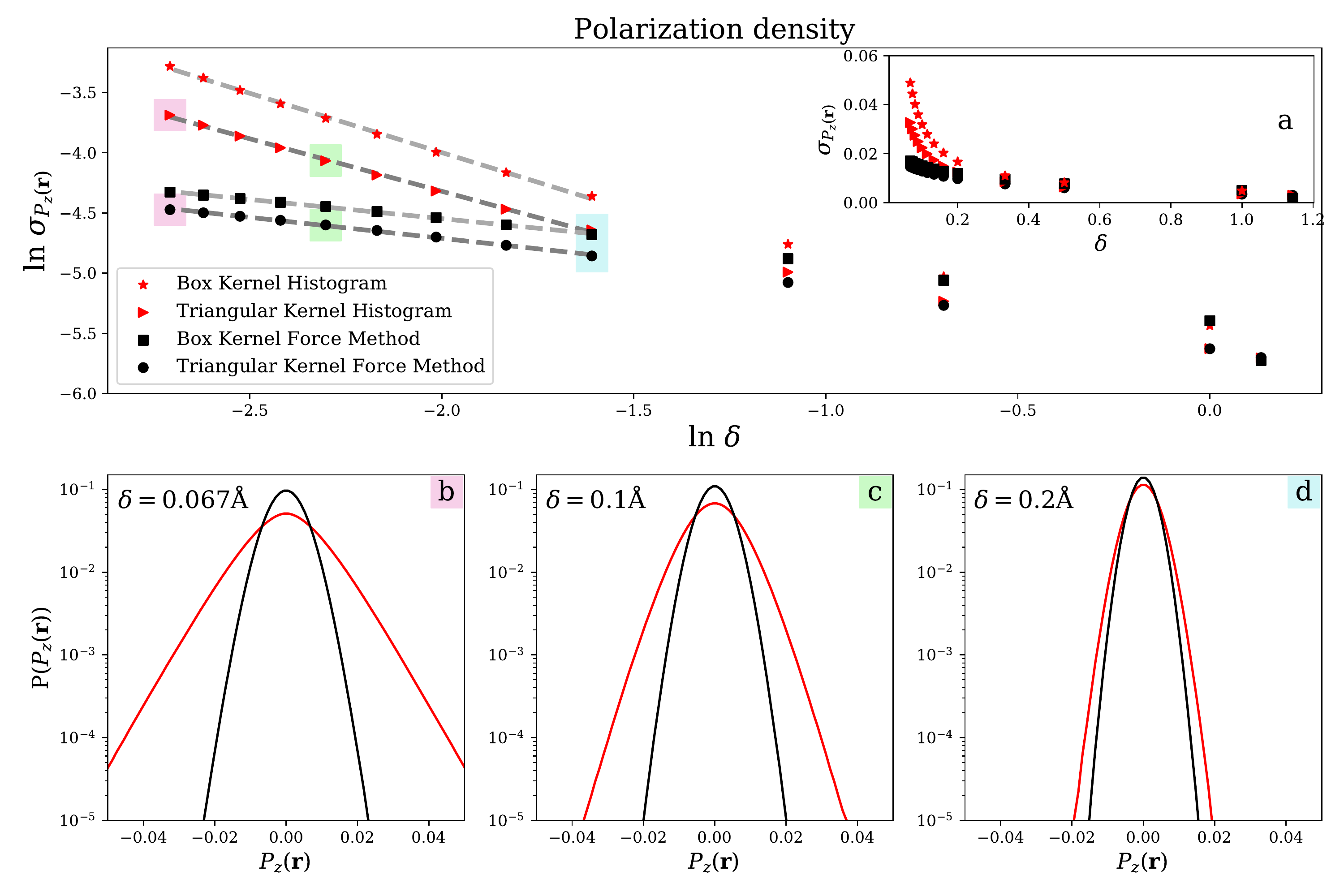}
  \caption{Panel a shows the effect of changing $\delta$ on the standard
deviation of the polarization densities for the histogram method(red stars for
box kernel grids; red triangles for triangular kernel grids), and the force
method(black squares for box kernel grids; black circles for triangular kernel
grids), shown on a log-log scale in the main figure and on a linear scale in the
inset. Panels b through d show the distributions of densities obtained for three
delta values $\delta =$ 0.067\AA , 0.1\AA , and 0.2\AA. The resulting
distributions of $P_{z}(\bfr)$ for the histogram method (red) and the force
method (black). The data points on panel a which correspond to the data in
panels b through d are shown by colored squares. Dashed lines are linear fit to the data in the region from $\delta=0.2$ to $\delta=0.067$, with the slopes of these lines reported in Table\,1.}
  \label{fig:noisepol}
\end{figure*}
\end{center}

We now turn to the distribution of the polarization density, illustrated in
Fig.\,\ref{fig:noisepol}. As for number density, the standard deviation
of polarization density is found to be similar for the two methods at larger
values of $\delta$ but is shown to diverge with decreasing grid spacing. In
panels b to d, we see again an increasing width of distributions with decreasing
$\delta$, and that the distributions for the force-based method
remain largely Gaussian. For the histogram-based
method, however, the form of the distribution becomes increasingly less Gaussian
with increasing $\delta$.
 
\begin{table}[]
\caption{
\label{table:slopes}
The calculated slopes for the log-log plots of the standard deviation of densities against $\delta$ in Figs.\,\ref{fig:noisenorm} and\,\ref{fig:noisepol}. }
\begin{tabular}{|l|l|l|}
\hline
Method                         &  $\rho(\bfr)$     &  $P_z(\bfr)$     \\ \hline
Box Kernel Histogram           & -0.88 & -0.87 \\ \hline
Triangular Kernel Histogram    & -0.98 & -0.98 \\ \hline
Box Kernel Force Method        & -0.33 & -0.32 \\ \hline
Triangular Kernel Force Method & -0.37 & -0.35 \\ \hline
\end{tabular}
\end{table}

Figures~\ref{fig:noisenorm} and~\ref{fig:noisepol} further demonstrate that for both densities, using the
triangular kernel to compute the force or number density on the grid is superior
to the box kernel (or conventional histogram), owing to the lower overall
standard deviation for all values of $\delta$. Furthermore, the improvement
gained by
changing to a triangular kernel is greater for the histogram method
than for the force method. However, it should be noted, as can be seen from
Table 1 , that the standard deviation increases less with decreasing $\delta$
for the box kernel than for the triangular kernel for each method and density.
From this point on, the more effective triangular kernel will be used, however it must be noted that this kernel decreases the improvement resulting from the new force method relative to the box kernel. Detailed results for the box kernel for the solvated water system are provided in the supplementary information.

\subsection*{Solvation: water around water}
\begin{center}
\begin{figure}[t!]
  \includegraphics[width=4.5cm]{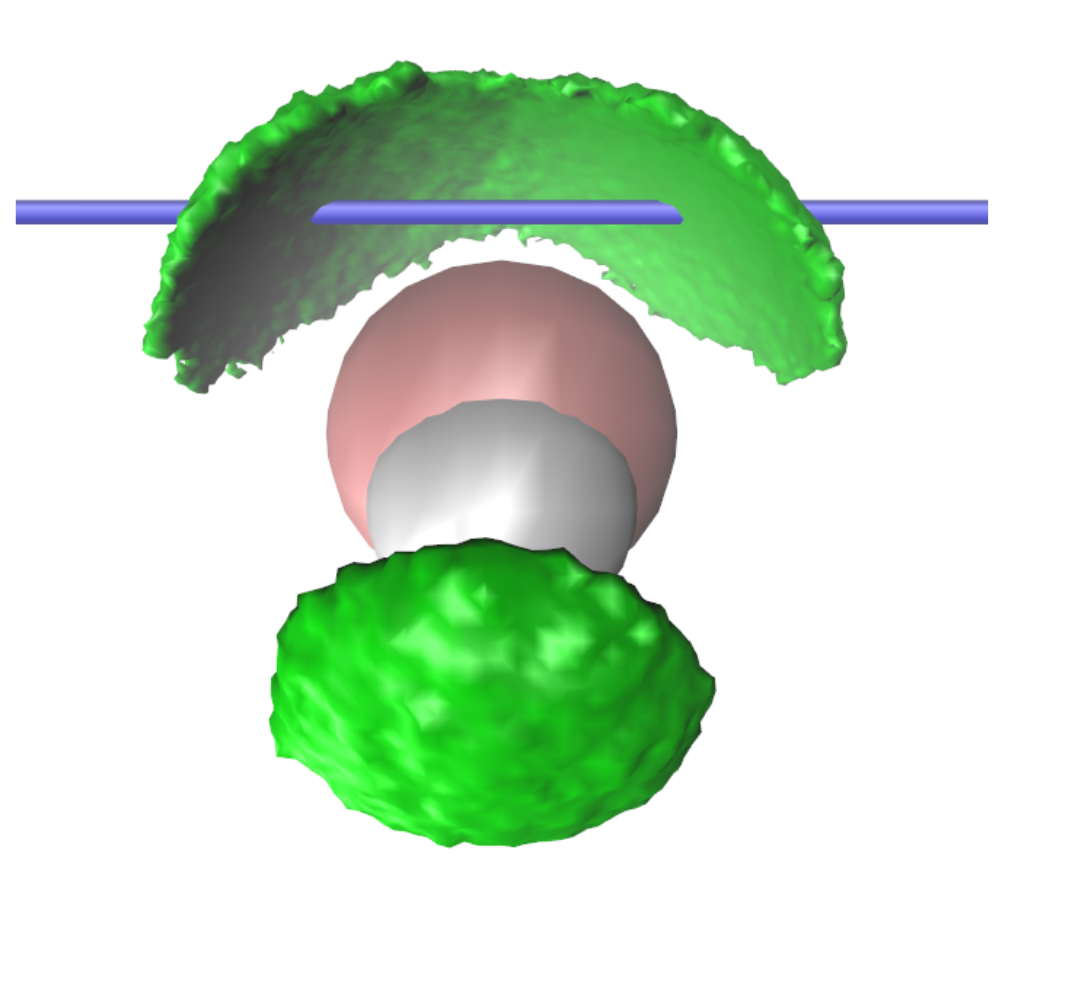}
  \caption{A rendered image showing the location of the characterization trace used for the
1D-analysis of number and polarization densities. 
In the case of polarization the $z$ axis is taken to run from the left to the
right of this image. The isosurface corresponds to a number density
$\rho(\bfr)=0.07$\AA$^{-3}$ and illustrates the position of water molecules in the first solvation shell.}
  \label{fig:traceExp}
\end{figure}
\end{center}
As previously mentioned, in order to explore the efficacy of the methodology for
the extraction of 3D densities we consider first a simple model system
consisting of a frozen water molecule in bulk water. In the following we analyze
the 3D density along a single line of voxels in the direction ($z$)
perpendicular to the molecular plane cutting the latter along the dipolar axis
($y$) at a position 0.6\,\AA\ away from the oxygen atom as shown in
Fig.\,\ref{fig:traceExp}. The green isosurface defines the water molecules
solvation shell. The traces that are taken along this line of voxels
thus contain two peaks, and a void between them. When considering this plot from
a more chemical perspective, we notice that the isosurface plot highlights three main regions of high density. The first region sits above the oxygen atom and consists of two lobes, corresponding to molecules donating a hydrogen bond to the constrained water molecule's oxygen atom. The other two regions of high density correspond to hydrogen molecules receiving a hydrogen bond from the central molecule.

In the following, we analyze qualitatively the effect of grid spacing and
trajectory length on the accuracy of the histogram and force methods, for the
number density. We then proceed with polarization density before qualitatively
examining the benefits of the force method for 3D representations of both types of density.

\subsubsection{Number density: effect of grid spacing}

\begin{figure*}
\includegraphics[width=\linewidth]{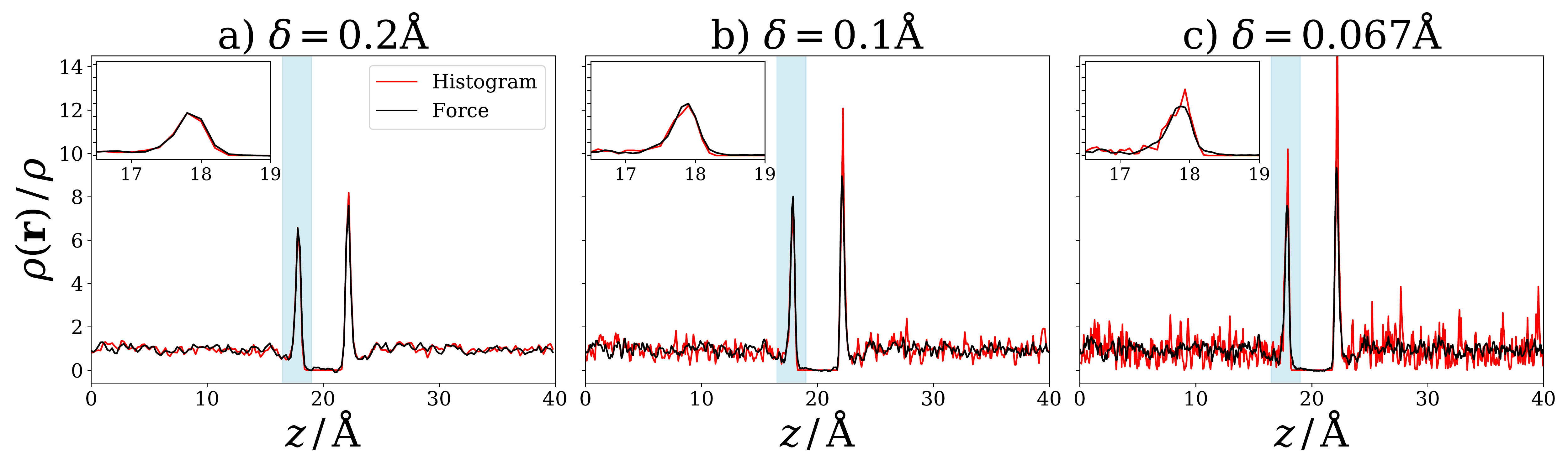}
  \caption{Number density extracted by means of the traditional histogram method
(red) and the force method (black), for a single line of voxels as shown in
Fig.\,\ref{fig:traceExp}. The data was extracted over the course of 4\,ns with
snapshots taken every 50\,fs. This figure specifically explores the effect of
varying grid spacing, $\delta=0.2, 0.1$ and $0.067$\AA\ for panels a to c,
respectively. 
The insets with each figure are
close-ups of the rising edge highlighted in blue in the main figure.}
  \label{fig:DensSizeSmall}
\end{figure*}

Fig.\,\ref{fig:DensSizeSmall} shows the 3D number density along the trace
illustrated in Fig.\,\ref{fig:traceExp} for the histogram and force methods
using grids with cubic voxels with edge length $\delta=$ 0.067\,\AA\ ,
0.1\,\r{A} , and 0.2\,\AA. The latter is typical of the precision  used in
classical density functional theory studies of aqueous
systems\,\cite{Ding2017,Zhao2011,Jeanmairet2013}. In all cases, the density is
reconstructed from the same configurations sampled every 50\,fs from the 4\,ns
trajectory. The inset in each panel shows the form of the rising edge
highlighted in blue in the main part of the panel. We first note that the basic
form of each trace is identical regardless of the method and of the grid size.
We observe the two main features in the form of the traces with a void centered
on $z\,=$\,20\,\AA\ and two peaks present at 17.8\AA\ and 22.2\,\AA. While for
the considered number of configurations the two methods provide comparable
results for the large voxels ($\delta$=0.2\,\AA), the noise increases much more
dramatically with decreasing $\delta$ in the case of the histogram method.
Importantly, while the results of the force method are qualitatively unchanged
when decreasing $\delta$, those of the histogram method develop undesirable
features such as a growing asymmetry between the two peaks, both of which are increasingly poorly described.

\subsubsection{Number density: effect of trajectory length}

The previous analysis of grid spacings demonstrates the benefit of
using the force method (compared to the histogram method) as the amount of data
per voxel decreases. In practice, one is often interested in sampling the
density for a fixed grid and the question to address is: how much data 
do we need to reach a given accuracy? We therefore examine, for a fixed grid size
$\delta=0.1$\,\AA, the effect of the quality of the available data by
considering several trajectory lengths (and sampling rates): an extremely long
simulation and one each with lengths representative of typical classical and ab initio MD simulations.

\begin{figure*}[t!]
\includegraphics[width=\linewidth]{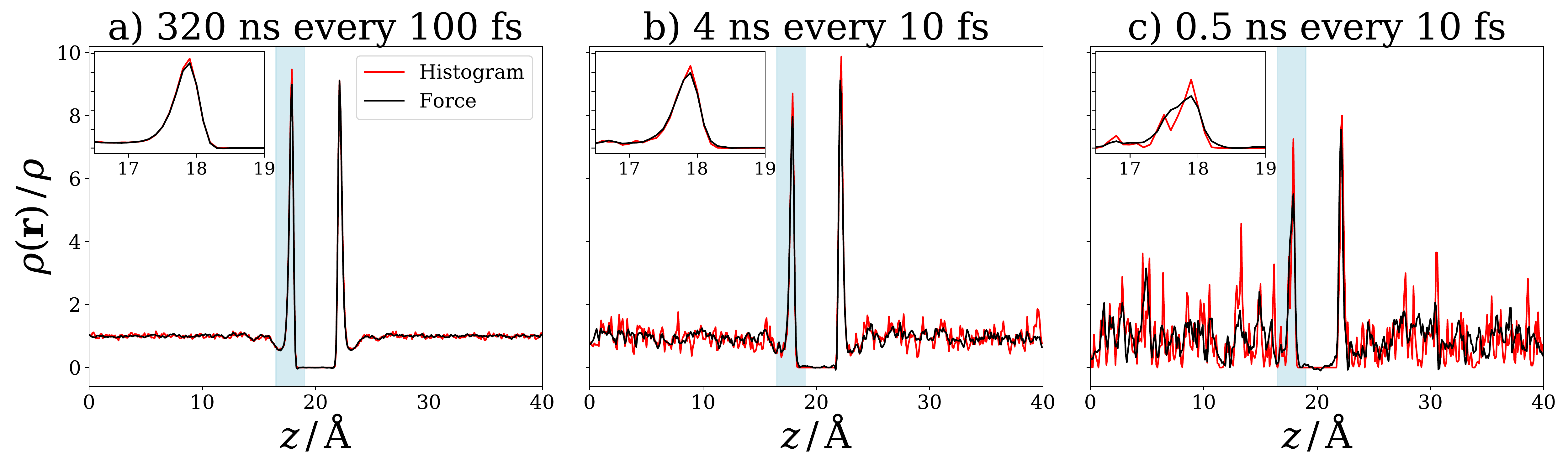}
  \caption{Number density extracted by means of the traditional histogram method
(red) and the force method (black), for the single line of voxels shown in
Fig.\,\ref{fig:traceExp}. The data was extracted on a grid with a spacing of
0.1\AA\ with snapshots taken every 10\,fs for the two shorter trajectories, and
every 100\,fs for the longer trajectory. This figure specifically explores the
effect of varying trajectory length: 320, 4 and 0.5~ns for panels a, b
and c, respectively. The insets are close-ups of the rising edges highlighted in blue in the main figures.}
  \label{fig:DensLengthSmall}
\end{figure*}

As a reference, we report in Fig.~\ref{fig:DensLengthSmall}a the results
for a set of \num{3e6} configurations (from a 300\,ns trajectory with samples
every 100\,fs). Even in that extreme case where data is
not scarce, the benefit of the force method
remains obvious. Decreasing the number of frames to \num{4e5} and increasing the
correlations between frames (4\,ns sampled every 10\,fs), and even \num{5e4}
(500\,ps sampled every 10\,fs) only makes this improvement more obvious.

\subsubsection{Polarization density: effect of grid spacing}

Thus far we have discussed only 3D number density (corresponding to $a_i=1$ in
Eq.~\ref{eq:aiINTRO}, and~\ref{eq:gradrhoa}) that we extracted from the
force density as in Borgis\textit{et al.}~\cite{borgis_computation_2013}. This is in itself novel as the methodology has not previously been applied to constrained molecules. We now turn to the other novel realization of this work, by considering a case $a_i \neq 1$, namely polarization ($a_i =\mathrm{p}_z$), i.e. the projection along the axis corresponding to the trace in Fig.\,\ref{fig:traceExp}.

\begin{figure*}[t!]
\includegraphics[width=\linewidth]{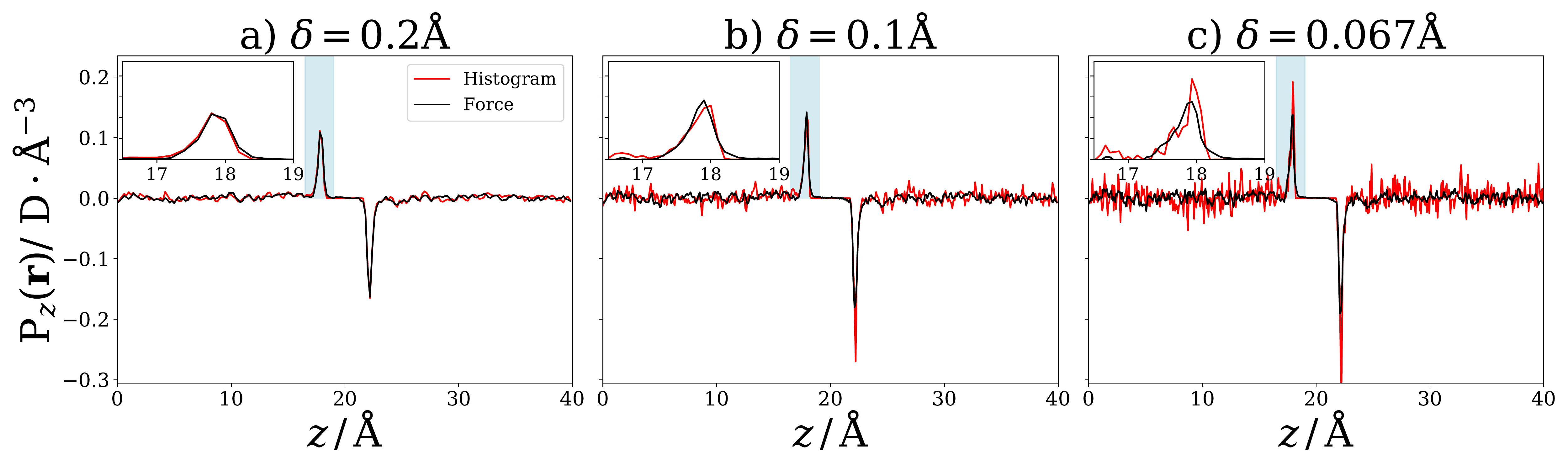}
  \caption{Polarization density extracted by means of the traditional histogram
method (red) and the force method (black), for the single line of voxels as
shown in Fig.\,\ref{fig:traceExp}. The data was extracted over the course of
4\,ns with snapshots taken every 50\,fs. This figure specifically explores the
effect of varying grid spacing, $\delta=0.2, 0.1$ and $0.067$\AA\ for panels a to c,
respectively. The insets are close-ups of the rising edge highlighted in blue on the main figures.}
  \label{fig:PolLengthSmall}
\end{figure*}

Fig.\,\ref{fig:PolLengthSmall} shows traces for the polarization density taken
along the path shown in Fig\,\ref{fig:traceExp}, with the axis running from left
to right. The figure has an identical layout to the previous cases with panels
showing decreasing $\delta$ running from panel a to c. In all cases, we see
positive polarization on approaching the
constrained water molecule and negative polarization as we move away from that
central water molecule. The plots should feature a center of inversion at point
(20,0) due to the orthogonal relationship between the constrained water
molecule's $\sigma_{v}$ plane of symmetry and the traces that are plotted. The
maintenance of this inversion symmetry will be a critical factor in accessing
the accuracy of the results of both methods.

As was the case for the number density, the two methods produce near-identical
results for the largest grid spacing ($\delta=$0.2\,\AA). We further note that
at this grid spacing both peaks are roughly symmetric with respect to the
expected center of inversion. However, as before, as we decrease the grid spacing (and thus the amount of data present per voxel) the force method is increasingly superior in extracting the polarization density. This is evidenced by the larger increase in the noise level with decreasing $\delta$ for the histogram method, as well as the poorer maintenance of the inversion symmetry in that case.

\subsubsection{Resolution of the 3D structure}

\begin{figure}[h!]
  \includegraphics[width=\linewidth]{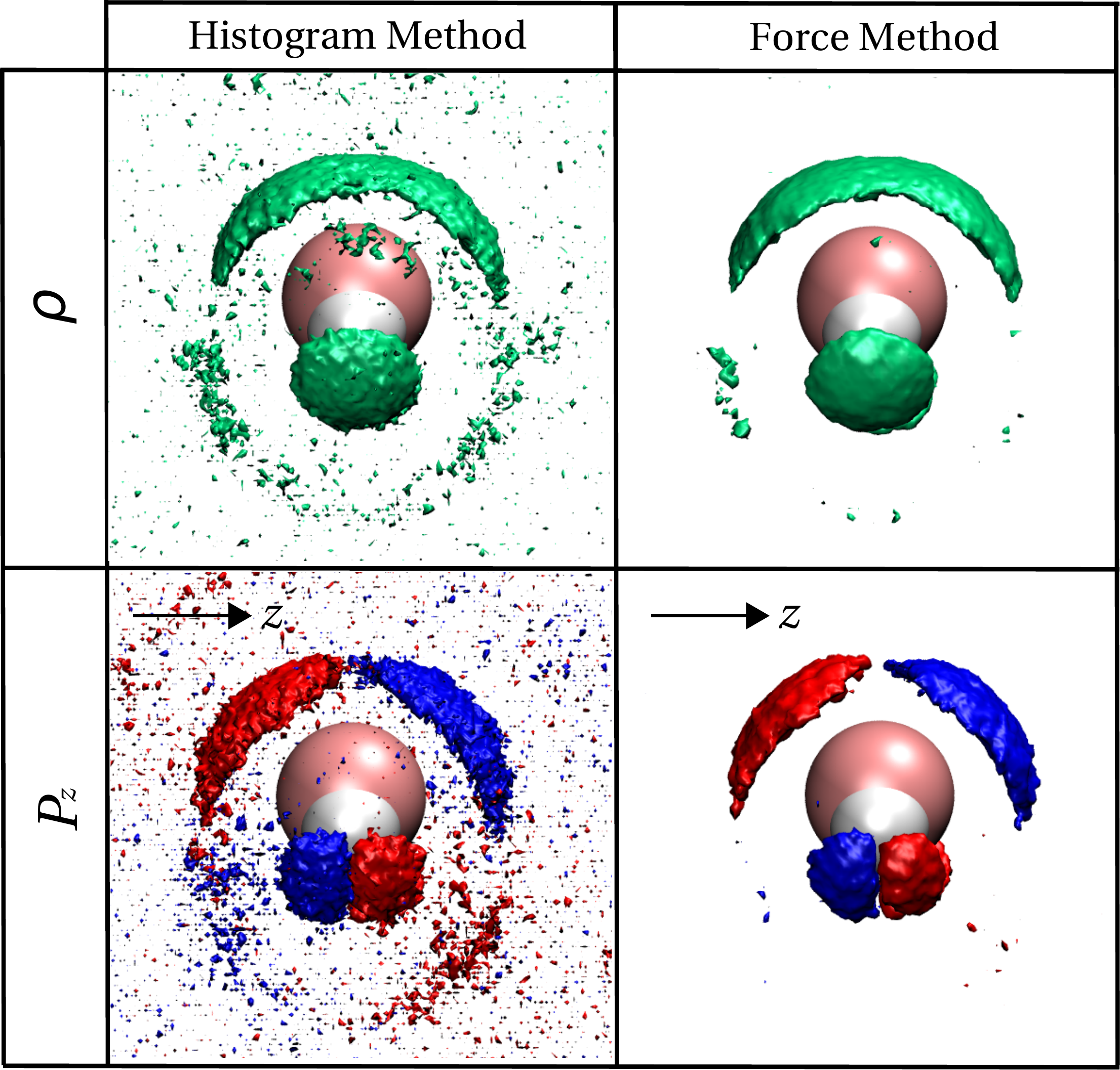}
  \caption{Isosurface plots for number density and polarization density for the two different methods of extraction of 3D densities. The green isosurface bounds the region where the number density is greater than 0.07\,\AA$^{-3}$ . For areas of polarization density blue (negative) and red (positive) bound less than -0.035\,D$\cdot$\AA$^{-3}$ and greater than +0.035\,D$\cdot$\AA$^{-3}$ respectively.}
  \label{fig:panelWater}
\end{figure}

Fig.\,\ref{fig:panelWater} brings the implications of the previously observed
trends into sharp relief. The isosurfaces are plotted for 3D densities for a
grid size $\delta=$ 0.1\,\AA. The top and bottom two panels in the figure
illustrate the 3D number and polarization density respectively extracted using
both methods. The force method provides a large improvement on the histogram
method in a number of key ways. As previously noted when looking at single
traces and the density distributions without a constrained molecule, the noise
away from the solvation shell is largely reduced with the new method, both for
number and polarization densities. Likewise, we see a massive improvement in the
resolution of the isosurfaces. This has large advantages for the further study and analysis of the 3D structure of the system, e.g. to identify basins of high density, a process that will become increasingly difficult with increasing roughness.

Fig.\,\ref{fig:panelWater} additionally illustrates a feature specific to the polarization,
namely a transition from positive polarization density to negative polarization
density when passing through the molecular plane. For the force method, this transition is abrupt and occurs clearly at the location of the $\sigma_v$ molecular plane. This is a clear improvement on the histogram method where this boundary is poorly defined.

\subsection{Lysozyme Solvation 4\,ns}
After the demonstration of the success of the force method on a somewhat
academic first case, we finally illustrate its relevance on a physically more
appealing example, namely the lysozyme protein. In contrast with the more
symmetric constrained water system, the globular protein studied here cannot be
well understood by studying the polarization density. Instead, we study here the
3D charge density, where the value of $a_i$ in Eq.\,\eqref{eq:aiINTRO} is taken
to be $q_i$. In line with the previously described equations for rigid
molecules, the position and charge of each atom is considered when calculating
the charge density, but the relevant force is the total force acting on the rigid molecule
(see Eqs.~\ref{eq:gradrhoa} and~\ref{eq:totforce}).

\begin{center}
\begin{figure}[h!]
  \includegraphics[width=\linewidth]{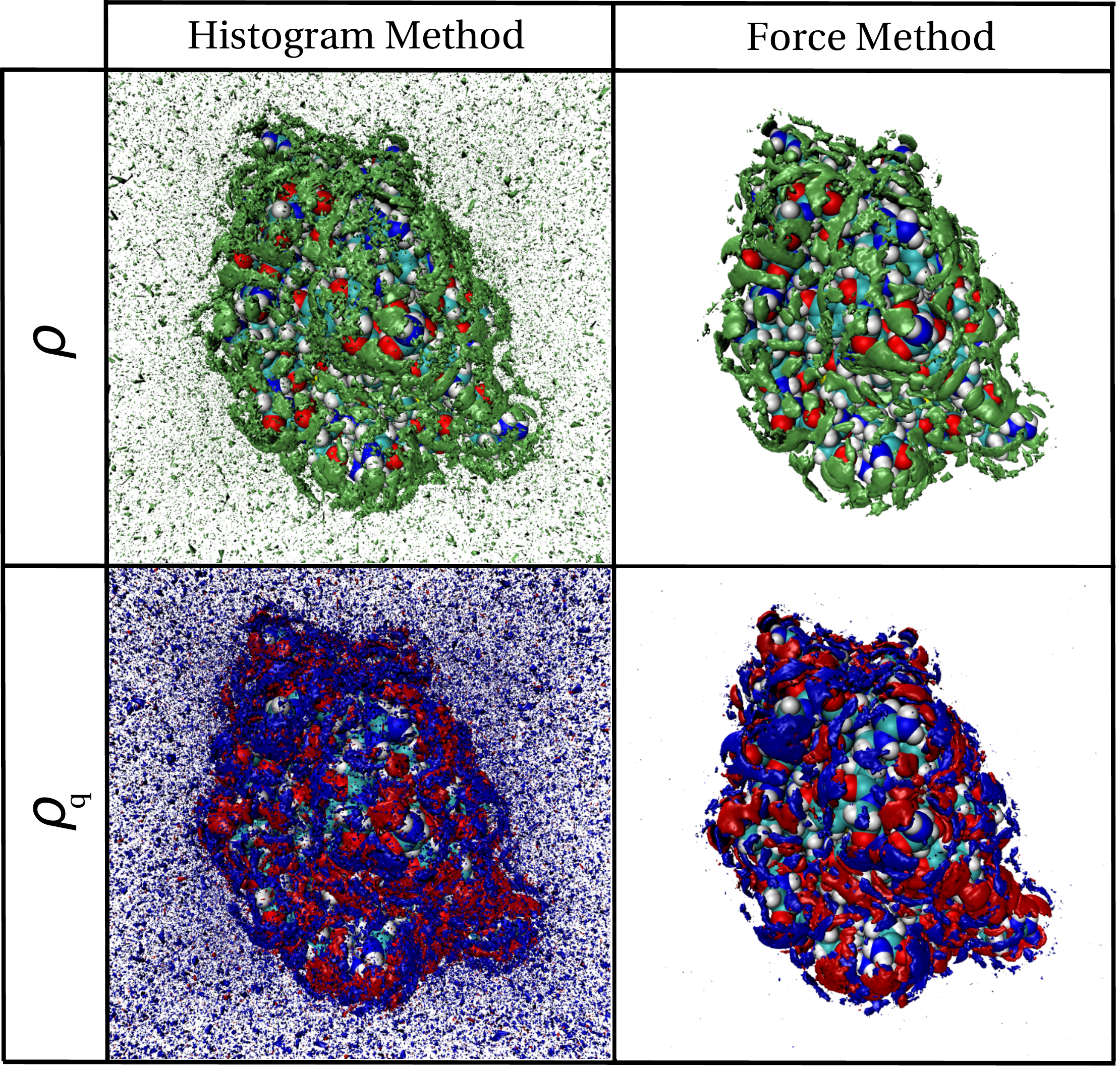}
  \caption{Isosurface plots for number density, $\rho$, and charge density,
$\rho_q$, for the two methods of extraction of 3D densities. The green
isosurfaces bound the regions where the number density greater than
0.1\,\AA$^{-3}$. The red (positive) and blue (negative) surfaces bound areas
where the magnitude of the charge density exceeds $\pm$0.1\,$e\cdot$\AA$^{-3}$.}
  \label{fig:panelLyso}
\end{figure}
\end{center}

The isosurfaces plots in Fig.\,\ref{fig:panelLyso} show the potential of the new
method in the understanding of solvation. Firstly, in both
cases, number and charge density, the histogram method leads to a large amount
of noise away from the protein molecules surface. Beyond making the image
less aesthetically pleasing this leads to two further problems in the
density extracted by the histogram method relative to the force method. This
noise is symptomatic of a poor baseline that will cause difficulties for any
subsequent analyses that need to be performed. Further to this, it causes a
large difficulty in resolving the shape of the solvation shell at its outer
boundary. In the case of the charge density, we see even better 
the improvements obtained thanks to the new method. These lobes of positive and
negative charge density allow one to identify the directionality of the
coordination by water molecules at each site on the protein. Overall these results show a massive
improvement in the resolution of the isosurfaces and show the clear applicability
of this methodology to real systems beyond the model systems intensively studied
in the previous sections.

\section{Conclusion}

We have presented a reduced variance method for the calculation of not only 3D
number densities but also other generic 3D densities by molecular simulations.
The data collection of the local force densities instead of the number densities
and their post-processing is as easy and inexpensive as in the conventional
density histograms collection, requiring only an extra 3D-FFT at the end to
transform from force densities to number densities. We have further extended
this method to the common case of rigid molecules described by distance
constraints, such as the popular SPC/E or TIPnP water models. This new force
density method appears more or less statistically equivalent to the conventional
method for voxel sizes above $\delta = 0.2$\,\AA, but is much more efficient
below that size. Furthermore, the variance of the results depends only slightly
on  $\delta$. This improved statistical efficiency makes it possible to reach a
given prescribed variance of the computed densities with shorter trajectories,
thus at a reduced simulation cost. We clearly illustrated for the water
structure around a small molecular solute (specifically water in water) as well
as for a complex molecular object (lysozyme protein), that for a given
simulation time the force method enables better resolution of the 3D structure,
including individualization of the density peaks and equalization of
the background and that it leads to a much clearer visualization
(figures 7 and 8). We thus believe that the method has already a wide range of
useful applications in many fields when it comes to characterizing by simulation the molecular solvation structure of complex biomolecular or solid-liquid interfaces. Several theoretical/technical challenges remain, such as a possible optimal mixing between density histograms and force histograms, or the expansion from force densities to force divergence densities evoked in Ref. \citenum{borgis_computation_2013}. Further work along these lines is underway.

\section*{Supplementary Material}

Versions of graphs contained within the water molecule solvation section are presented for the box kernel in the supplementary information.

\section*{acknowledgements}

SWC and BR acknowledge financial support from the Ville de Paris (Emergences,
project Blue Energy). The research of SWC leading to these results has received
funding from the People Programme (Marie Curie Actions) of the European Union's Seventh Framework Programme (FP7/2007-2013) under REA grant agreement n. PCOFUND-GA-2013-609102, through the PRESTIGE programme coordinated by Campus France. Finally we thank Maximilien Levesque  and Guillaume Jeanmairet for many useful discussions.

%

\widetext
\clearpage
\begin{center}
\textbf{\Large Supporting information: Computing three-dimensional densities from force densities improves statistical efficiency}
\end{center}
\setcounter{equation}{0}
\setcounter{figure}{0}
\setcounter{table}{0}
\setcounter{page}{1}
\makeatletter
\renewcommand{\theequation}{S\arabic{equation}}
\renewcommand{\thefigure}{S\arabic{figure}}
\renewcommand{\thepage}{S\arabic{page}}
\renewcommand{\bibnumfmt}[1]{[S#1]}
\renewcommand{\citenumfont}[1]{S#1}

\maketitle
\onecolumngrid
\section*{Water number density around water: results using a box kernel}

\begin{figure}[b!]
\includegraphics[width=\linewidth]{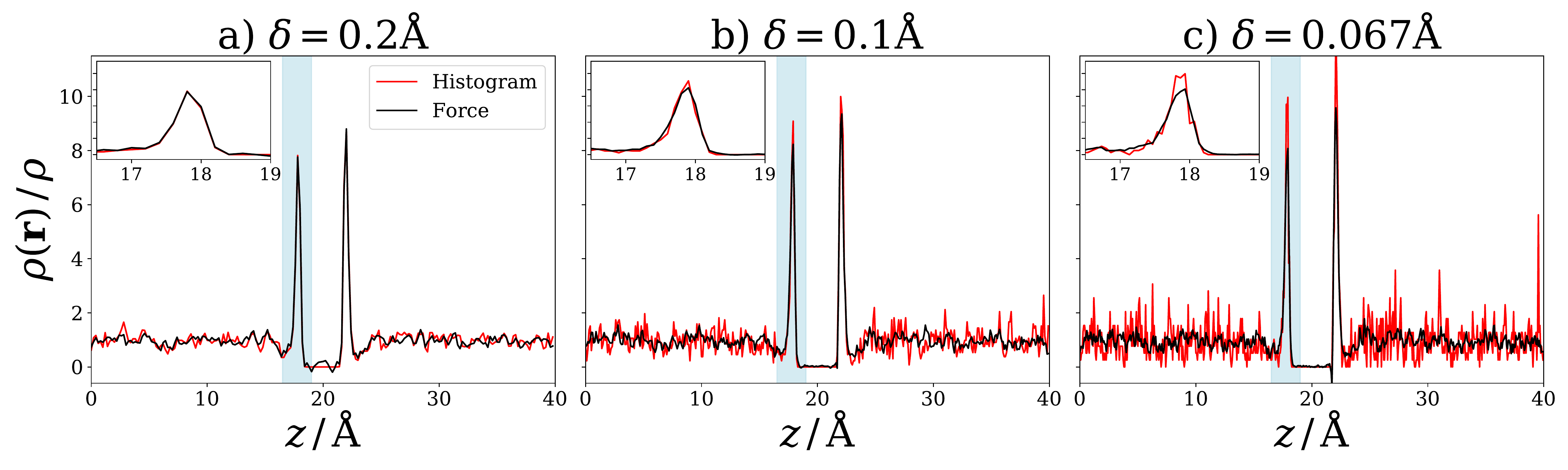}
  \caption{Number density extracted by means of the traditional histogram method (red) and the force method (black), for a single line of pixels as shown in Fig.\,3 of the main paper. The data was extracted over the course of 4\,ns with snapshots taken every 50\,fs. This figure specifically explores the effect of varying grid spacing, $\delta=0.2, 0.1$ and $0.067$\AA\ for panels a to c,
respectively. The insets within each figure are close ups of the rising edge highlighted in blue in the main figure. In contrast with Fig.\,4 in the main paper a box kernel was used to extract the grid in this example.}
  \label{fig:nn}
\end{figure}

In the main paper a triangular kernel was used to obtain the grids of number
density, polarization density, force density, and the polarization density
equivalent of the force density. The triangular kernel was employed in order to
give the histogram method the best possible chance against the force method, as
it significantly minimizes the effects of digitization on the resulting
histogram. However when trying to obtain a 3 dimensional density one might
naively prefer to use a box kernel, which is the conventional way of taking 3D
histograms. In Fig.~\ref{fig:nn} we present the results obtained for the number density around a solvated water molecule obtain using a box kernel (this figure is identical to Fig.\,4 of the main paper but for the change of kernel). The new force method provides the ability for reduced variance extraction of number densities particularly at lower values of $\delta$. One additional effect of the application of the force method to box kernel grids is that it removes the digitization of the data which is observed for the direct binning of positions. This is in accord with the analysis of density fluctuations in three dimensional space, which showed decreased standard deviation with the triangular kernel for exactly this reason.

\begin{figure}[t!]
\includegraphics[width=\linewidth]{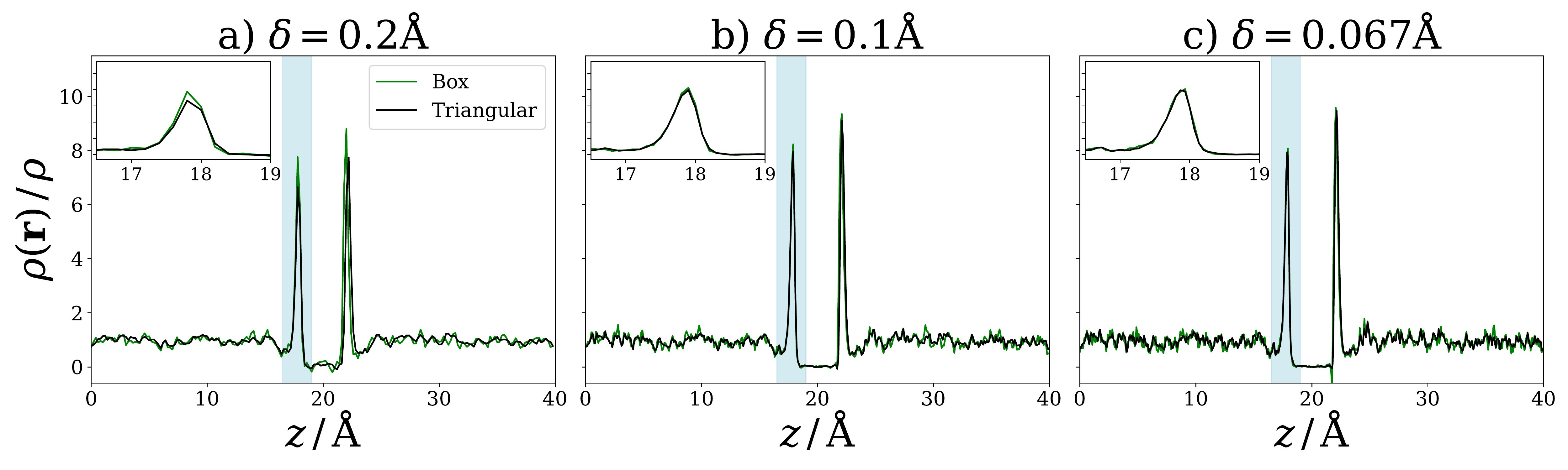}
  \caption{Number density extracted by means of the force method for grids created using triangular (black), and box (green) kernels for a single line of pixels as shown in Fig.\,3 of the main paper. The data was extracted over the course of 4\,ns with snapshots taken every 50\,fs. This figure specifically explores the effect of varying grid spacing, $\delta=0.2, 0.1$ and $0.067$\AA\ for panels a to c,
respectively. The insets within each figure are close ups of the rising edge highlighted in blue in the main figure.}
  \label{fig:nnTRI}
\end{figure}

We now move on to compare the results of the force method using both triangular
and box kernels displayed in Fig.~\ref{fig:nnTRI}. In the main paper the analysis of the density fluctuation showed a slightly smaller noise level for the triangular kernel. The difference in noise across all values of $\delta$ for the two methods is however not the most obvious difference, it is more apparent that the grids formed using a triangular kernel provide better resolution of the edge of the void.

\section*{Water polarization density around water: results using a box kernel}

\begin{figure}[b!]
\includegraphics[width=\linewidth]{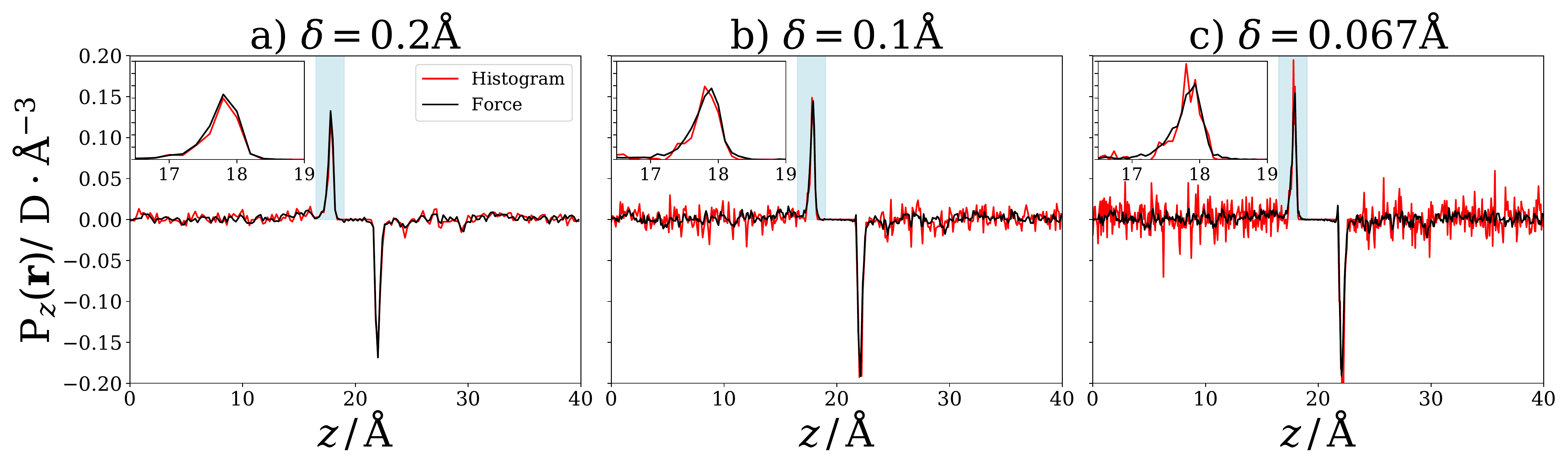}
  \caption{Polarization densities extracted by means of the traditional histogram method (red) and the force method (black). The polarization density is shown for a single line of pixels as shown in Fig.\,3 of the main paper. The data was extracted over the course of 4\,ns with snapshots taken every 50\,fs. This figure specifically explores the effect of varying grid spacing, $\delta=0.2, 0.1$ and $0.067$\AA\ for panels a to c,
respectively. The insets within each figure are close ups of the rising edge highlighted in blue in the main figure. In contrast with Fig.\,6 in the main paper a box kernel was used to obtain the grid in this example.}
  \label{fig:nnpol}
\end{figure}

\begin{figure}[t!]
\includegraphics[width=\linewidth]{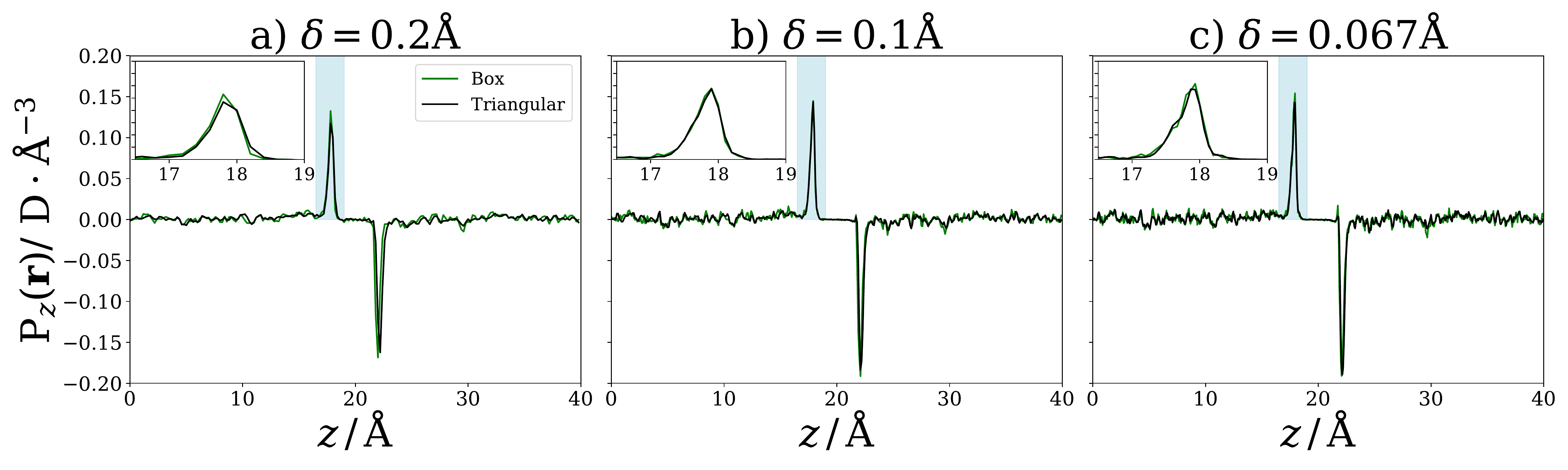}
  \caption{Polarization density extracted by means of the force method for grids created using triangular (black), and box (green) kernels for a single line of pixels as shown in Fig.\,3 of the main paper. The data was extracted over the course of 4\,ns with snapshots taken every 50\,fs. This figure specifically explores the effect of varying grid spacing, $\delta=0.2, 0.1$ and $0.067$\AA\ for panels a to c,
respectively. The insets within each figure are close ups of the rising edge highlighted in blue in the main figure.}
  \label{fig:nnTRIpol}
\end{figure}

Fig.~\ref{fig:nnpol} shows the traces for the polarization density, using both the histogram
and force methods, using grids obtained with the box kernel. In this example we
see, as in the main paper and the number density, that the form of the traces is
roughly the same for the two methods. The amount of noise is greater for the
traditional method than for the force method for the two finer grids, but, as
before, is similar for $\delta= 0.2$\,\AA. As in the case of number density, the
difference in the ability to resolve the void becomes even more apparent on
studying Fig.~\ref{fig:nnTRIpol} which shows the results obtained with force method for the two kernels. A distinct fish hook feature is observed on the second rising edge for $\delta=$0.067\,\AA\ for the box kernel which isn't present for the triangular kernel.

\end{document}